\documentclass[12pt,preprint]{aastex}

\def\LIR{L_{\rm IR}}
\def\Lsun{L_\odot}

\def\Mpc{\,{\rm Mpc}}

\newcommand{\SFR}{{\rm SFR}}

\newcommand{\myear}{M_\odot\, {\rm yr^{-1}}}

\begin{document}

\title{THE ROLE OF MAJOR GAS-RICH MERGERS ON THE EVOLUTION OF GALAXIES FROM THE BLUE CLOUD TO THE RED
SEQUENCE}
\author{Rui Guo\altaffilmark{1, 2, 3},
Cai-Na Hao\altaffilmark{3},
X. Y. Xia\altaffilmark{3},
Shude Mao\altaffilmark{4,2,5},
Yong Shi\altaffilmark{6,7}
}
\altaffiltext{1}{National Astronomical Observatories, Chinese Academy of Sciences, 20A Datun Road, Chaoyang District, Beijing 100012, China}
\altaffiltext{2}{University of Chinese Academy of Sciences, Beijing 100049, China}
\altaffiltext{3}{Tianjin Astrophysics Center, Tianjin Normal University, Tianjin 300387, China; E-mail: cainahao@gmail.com}
\altaffiltext{4}{Physics Department and Tsinghua Center for Astrophysics, Tsinghua University, Beijing, 100084, China}
\altaffiltext{5}{Jodrell Bank Centre for Astrophysics, University of Manchester, Alan Turing Building, Manchester M13 9PL, UK}
\altaffiltext{6}{School of Astronomy and Space Science, Nanjing University, Nanjing 210093, China}
\altaffiltext{7}{Key Laboratory of Modern Astronomy and Astrophysics, Nanjing University, Ministry of Education, Nanjing 210093, China
}
\date{ }

\begin{abstract}

With the aim of exploring the fast evolutionary path from the blue cloud of
star-forming galaxies to the red sequence of quiescent galaxies in the local
universe, we select a local advanced merging infrared luminous and
ultraluminous galaxy (adv-merger (U)LIRGs) sample and perform careful dust
extinction corrections to investigate their positions in the SFR-$M_{\ast}$,
{\em u-r} and {\em NUV-r} color-mass diagrams. The sample consists of 89
(U)LIRGs at the late merger stage, obtained from cross-correlating the {\em
IRAS} Point Source Catalog Redshift Survey and 1 Jy ULIRGs samples with the
Sloan Digital Sky Survey DR7 database. Our results show that $74\%\pm 5\%$ of
adv-merger (U)LIRGs are localized above the $1\, \sigma$ line of the local
star-forming galaxy main sequence. We also find that all adv-merger (U)LIRGs
are more massive than and as blue as the blue cloud galaxies after corrections
of Galactic and internal dust extinctions, with $95\%\pm 2\%$ and $81\%\pm 4\%$
of them outside the blue cloud on the {\em u-r} and {\em NUV-r} color-mass
diagrams, respectively. These results, combined with the short timescale for 
exhausting the molecular gas reservoir in adv-merger (U)LIRGs ($3\times 10^{7}$ 
to $3\times 10^{8}$ years), imply that the adv-merger (U)LIRGs 
are likely at the starting point of the fast evolutionary track previously proposed by several groups.
While the number density of adv-merger (U)LIRGs is only $\sim 0.1\%$ of the blue cloud
star-forming galaxies in the local universe, this evolutionary track may play a
more important role at high redshift.

\end{abstract}


\section{INTRODUCTION}

Galaxies show a striking color bimodality in the color-magnitude and the
color-stellar mass (color-mass, hereafter) diagrams, both locally, based
on the Sloan Digital Sky Survey (SDSS) (Kauffmann et al. 2003b; Baldry et al.
2004, 2006), and at high redshift (Bell et al. 2004; Faber et al.
2007; Ilbert et al. 2010). Since the discovery of the bimodality,
several evolutionary pathways have been proposed for how the blue cloud
of star-forming galaxies evolves to the red sequence of quiescent galaxies
through the transition phase of the so-called green valley (e.g. Bell et
al. 2004; Faber et al. 2007; Marchesini et al. 2014; Schawinski et al.
2014).

There is also a tight correlation between the star formation rate
(SFR) and the stellar mass for normal star-forming galaxies, which is
referred to as the ``main sequence'' (MS) and changes with redshift in its
normalization out to redshift 6 (e.g. Brinchmann et al. 2004;
Elbaz et al. 2007; Noeske et al. 2007; Leitner 2012; Magdis et al. 2010;
Bouwens et al. 2012; Salmon et al. 2015). 
Meanwhile, starburst galaxies are located above the MS and are likely driven 
by gas-rich major mergers (Rodighiero et al. 2011; Hung et al. 2013).   

Furthermore, based on CO observations for galaxies at both low and high
redshifts, Daddi et al. (2010) and Genzel et al. (2010) presented evidence for
the bimodal behavior of star formation processes. Disk galaxies and
starburst galaxies, triggered by gas-rich major mergers, occupy distinct regions
in the SFR and gas mass diagrams. They suggested two star formation modes: a
long-lasting star formation mode where the gas reservoir is consumed slowly for
disk galaxies, and a rapid star formation mode for starburst galaxies which
have high star formation efficiencies ($\sim$ 10 times higher than those of disk
galaxies) due to their much higher dense gas fractions (Gao \& Solomon 2004). It
is likely that there exists a close link between the two star formation
modes and the different evolutionary pathways. 

In recent years, there are many works investigating the bimodality of galaxies 
in the color-magnitude diagram, particularly on how the blue cloud star-forming galaxies are quenched
and evolve to the red sequence (quiescent) galaxies with a broad stellar mass
distribution. This is especially the case for the formation of massive quiescent galaxies.
From the results of DEEP2 and COMBO-17 surveys on the luminosity functions of blue and red galaxies to redshift 1, 
Faber et al. (2007) first proposed a mixed scenario for the formation of massive red quiescent galaxies.
They suggested that after star formation in the blue cloud galaxies 
is quenched, these galaxies leave the blue cloud and migrate to the red sequence, then they evolve slowly along
the red sequence through a series of dry mergers to form local massive red sequence galaxies. 
On the other hand, observations of high-redshift galaxies (van Dokkum et al. 2008;
Kriek et al. 2009) found that massive quiescent galaxies are substantially
more compact than local galaxies with similar stellar masses. Galaxies with intense
and concentrated starbursts, triggered by gas-rich major mergers, are possible progenitors 
of massive compact quiescent galaxies (Wellons et al. 2015). 
Actually, Barro et al. (2013) detected compact star-forming galaxies (cSFGs) at redshift 1.4-3 by 
combining high resolution HST/WFC3 images and multi-wavelength photometry of massive galaxies
($M_{\ast}\ge 10^{10} M_{\odot}$) in GOODS-S and UDS fields of
the Cosmic Assembly Near-infrared Deep Extragalactic Legacy Survey (CANDELS;
Grogin et al. 2011; Koekemoer et al. 2011). From the number densities, sizes, SFRs and dynamical properties 
of cSFGs, they are likely the progenitors of high-redshift compact quiescent
massive galaxies (cQGs, see also Barro et al. 2014a, 2014b). Then, Barro et al. (2013) proposed two evolutionary tracks
to form quiescent galaxies: the cSFGs formed by gas-rich processes are quenched rapidly and fade 
to compact massive early-type galaxies, subsequent minor dry mergers can puff up quiescent early-type galaxies
over time and grow in the quiescent phase; on the other hand, larger SFGs evolve slowly and become extended early-type galaxies directly.

More and more pieces of evidence support this two evolutionary pathway scenario, especially the fast track for the
formation of massive quiescent galaxies. Marchesini et al. (2014)
searched for the progenitors of local ultra-massive galaxies with $M_{\ast}>6\times 10^{11} M_{\odot}$ 
based on Ultra VISTA catalogs (Muzzin et al. 2013b) and concluded that the progenitors of red sequence
massive galaxies are red, massive and heavily dust-extincted starburst galaxies
and have never been on the blue cloud since redshift 3. It implies that the most massive
galaxies formed early and quickly, which has been confirmed by Mancini et al. (2015) based
on detailed investigation for 56 massive galaxies on an object-to-object basis in the GOODS-South field
at $1.4<z<2$. Combining the SDSS Data Release 7 (DR7; York et al. 2000; Abazajian et al. 2009) 
and the {\em Galaxy Evolution Explorer (GALEX)} databases (Martin et al. 2005), and
using the morphology of galaxies from the Galaxy Zoo project (Lintott et al.
2008; Lintott et al. 2011), Schawinski et al. (2014) found that the bimodality
of galaxies in the color-mass diagram changes dramatically, i.e., there is not a clear
green valley dip for both late-type and early-type galaxies. 
A further analysis for the green valley galaxies based on the UV-optical colors shows that for the
majority of late-type galaxies, their star formation rates decline
secularly over several Gyr. In contrast, only a small fraction of blue galaxies evolve rapidly 
to the red sequence by gas-rich mergers and transform from disk galaxies to spheroids that should be 
the recently quenched massive blue galaxies found by McIntosh et al. (2014)
based also on SDSS database. Most recently, Belli et al. (2015) 
investigated 62 massive galaxies at the redshift range $1<z<1.6$ based on Keck LRIS spectra 
and confirmed the fast quenching mode that is through post-starburst phase.
Actually, Yesuf et al. (2014) analyzed a sample of post-starburst galaxies and concluded
that the quenching mechanisms could be broadly classified as fast and slow tracks.
The rapid quenching (the fast track) in the local universe is triggered by merger-induced starbursts.
Therefore, it is worth studying the fast quenching process in detail 
from a local gas-rich major merging galaxy sample.

In fact, such a sample is already known: almost all local ultraluminous infrared galaxies 
(ULIRGs; $\LIR$\footnote{$\LIR$ is the integrated infrared luminosity 
between 8-1000 $\mu$m.}$>10^{12}\Lsun$) and a large fraction of luminous
infrared galaxies (LIRGs; $10^{11}\Lsun<\LIR<10^{12}\Lsun$) are gas-rich
interacting/merging galaxies (e.g. Sanders \& Mirabel 1996; Wang et al. 2006; Kaviraj 2009).
Such gas-rich major mergers funnel cold molecular gas reservoirs into the centers, 
leading to massive starbursts (Downes \& Solomon 1998).
Their stellar masses are moderate (Dasyra et al. 2006a, 2006b; da Cunha et al. 2010;
also see Section 4.2), but are higher than those of blue cloud
galaxies ($M_{\ast}< 3\times10^{10} M_{\odot}$) (Kauffmann et al. 2004).
Similarly, a large fraction of high-redshift ULIRGs are interacting/merging
and starburst galaxies (Kartaltepe et al. 2010, 2012; Zamojski et al. 2011; Soto \& Martin 2012;
Hung et al. 2014; Yan et al. 2014). Furthermore, Casey et al. (2014) found that
the dusty star-forming galaxies with SFRs above $50\, M_\odot\, {\rm yr^{-1}}$
(LIRGs and ULIRGs) are blue at both low and high redshifts and become bluer as their infrared
luminosity increases. Since LIRGs and ULIRGs are blue and have higher stellar masses 
than the blue cloud disk galaxies, these galaxies should occupy the bottom right corner of the color-mass diagram 
and may be the progenitors of massive red sequence galaxies. 
In fact, Chen et al. (2010) have investigated the
positions of 52 local ULIRGs in the {\em g-r} color-magnitude diagram without
applying internal extinction corrections, and found that $\sim$46\% ULIRGs lie
outside the 90\% level number density contour, with a few in the green valley or even on the red
sequence. Kilerci Eser et al.  (2014) obtained similar results with a
larger sample of 82 local ULIRGs. If the galaxy internal extinction
is corrected, these local ULIRGs would appear even bluer and brighter, possibly
occupying different locations in the color-magnitude and color-mass diagrams. For this purpose, we
select a local (U)LIRG sample and focus on those with advanced merging (adv-merger, hereafter) morphologies,
which are galaxies at the late gas-rich major merger stage with a single nucleus.
Furthermore, we perform the galaxy internal extinction
correction carefully and investigate how their locations change in
the SFR-mass and color-mass diagrams. 

The structure of this paper is as follows. In Sections 2 and 3 we describe the
sample selection and parameters estimation for our sample adv-mergers and
control samples.  In Section 4, we present our main results. We discuss and
summarize in Sections 5 and 6 respectively. Throughout this paper we adopt the Kroupa
(2001) initial mass function and a cosmological model of ${H}_{\rm
0}=70\,{\rm km \, s^{-1}\, Mpc^{-1}}$, $\Omega_{\rm m}=0.3$ and $\Omega_{\rm
\Lambda}=0.7$.

\section{SAMPLE SELECTION}

\subsection{Advanced Mergers}

We morphologically select advanced mergers from a SDSS {\em r}-band
limited sample of LIRGs and ULIRGs, most of which are in the redshift range of 
$z < 0.1$ and $0.1 < z < 0.25$, respectively. Specifically,
our dusty adv-merger sample is mainly drawn from the cross-correlation between 
the spectroscopic catalog of SDSS DR7 (York et al. 2000; Abazajian et al.
2009) and the {\em Infrared Astronomical Satellite (IRAS)} Point Source Catalog Redshift 
Survey (PSCz, Saunders et al. 2000), during which we used a 5$\arcsec$ searching radius. 
Given that ULIRGs are very rare in the local universe ($z<0.3$)\footnote{There is only one ULIRGs
within redshift 0.03 (Sanders \& Mirabel, 1996).} but most of them are in the late merging stage 
(e.g. Sanders \& Mirabel, 1996; Veilleux et al. 2002), 
the ULIRGs in our sample are mainly collected by searching for counterparts of the {\em IRAS} 1 Jy
ULIRGs sample (Kim \& Sanders 1998) in the SDSS DR7 photometric catalog. We describe the sample selection
procedures in more detail below.

Since reliable morphological classifications can only be achieved for SDSS
galaxies brighter than 15.9 mag in the {\em r}-band (Fukugita et al. 2004) and
the spectroscopic observations are incomplete for SDSS galaxies with {\em
r}-band magnitude brighter than 14.5 mag (Kauffmann et al.  2003b), we
restricted our infrared luminous and ultraluminous galaxy sample to a magnitude range of
$14.5<r<15.9$ after corrections for foreground Galactic extinction (Schlegel et
al. 1998). Within this magnitude range, 259 LIRGs and ULIRGs were obtained by the
cross-correlation between the SDSS DR7 and PSCz. We will discuss possible
selection effects caused by this magnitude cut in Section 4.2.1. Furthermore, we examined the
positions of SDSS fibers on the images for these galaxies, and found that 
for 6 galaxies the fiber spectra were obtained for off-center regions.
Considering that the SFR might be under-estimated by the H$\alpha$
emission line for these galaxies, we removed them from our sample.  This step
reduced the number of galaxies to 253, out of which there are only 3 ULIRGs.
Therefore, we perform further cross-correlation between
the photometric catalog of SDSS DR7 and the {\em IRAS} 1 Jy ULIRGs sample, 
yielding 63 ULIRGs, among which there are 39 objects with SDSS spectra.
As the morphologies of these ULIRGs have been well studied in the literature,
we did not put constraints on their {\em r}-band magnitudes. In total, the
sample consists of 314 LIRGs and ULIRGs after excluding 2 overlapping objects
in the above two cross-correlation procedures.

\begin{figure*}
\plotone{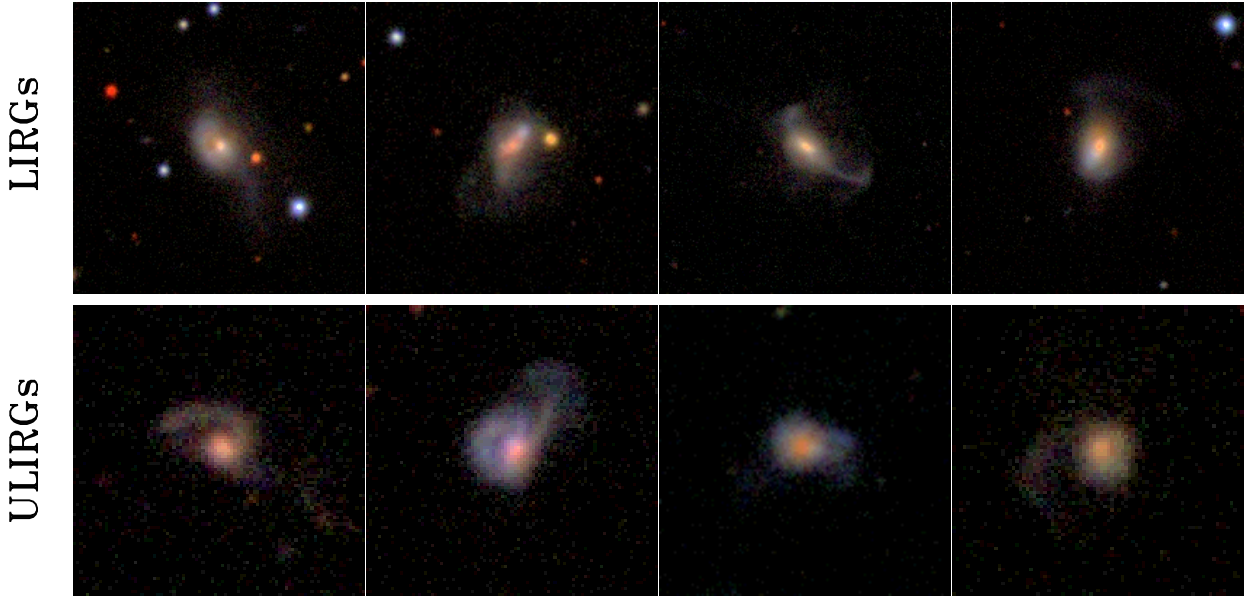}
\caption{True color images of 8 example adv-mergers with tidal tails or other
merging signatures, constructed from SDSS
({\em g, r, i}) images following Lupton et al. (2004). The top row shows 4
LIRGs and the bottom row shows 4 ULIRGs. The physical size of
each image is 80 $\times$ 80 $\rm kpc^2$.}
\label{colorim.eps}
\end{figure*}

Considering that we are only concerned with the adv-mergers in this work, we
further performed morphological classifications for the above 314 galaxies
visually. The classification scheme is similar to that of Wang
et al. (2006). The visual classifications are primarily based on their {\em
r}-band images from SDSS DR7 or {\em K}-band images from the 2.2 m telescope of
University of Hawaii (Kim et al. 2002). Then we used SDSS composite true color
images by combining {\em g, r, i} filter data as cross-check (Lupton et al.
2004).  The sample galaxies are classified into five classes: early-type like galaxies with
relaxed morphologies (12), spiral galaxies with bars or without bars (118), interacting 
galaxies with disturbed morphologies (55), pre-merging galaxies
with two or more nuclei (27) and advanced merging galaxies (adv-merger) 
with a single nucleus but some merger signs (100). Furthermore, we compared our
morphological classification results with the Galaxy
Zoo 1\footnote{ http://data.galaxyzoo.org/} (Lintott et al. 2008; Lintott et al.
2011).  There are just a few different classifications ($\sim$10\%), in which
some interacting or merging galaxies in our classifications were classified as spirals or
ellipticals in the Galaxy Zoo, but their images do show disturbed features.
Our classifications for early-type and spiral galaxies are almost the
same as those of Galaxy Zoo. We only select the 100 adv-merging galaxies as our working sample.
Figure \ref{colorim.eps} shows the true color images of 8 example adv-mergers.

\begin{figure*}
\plotone{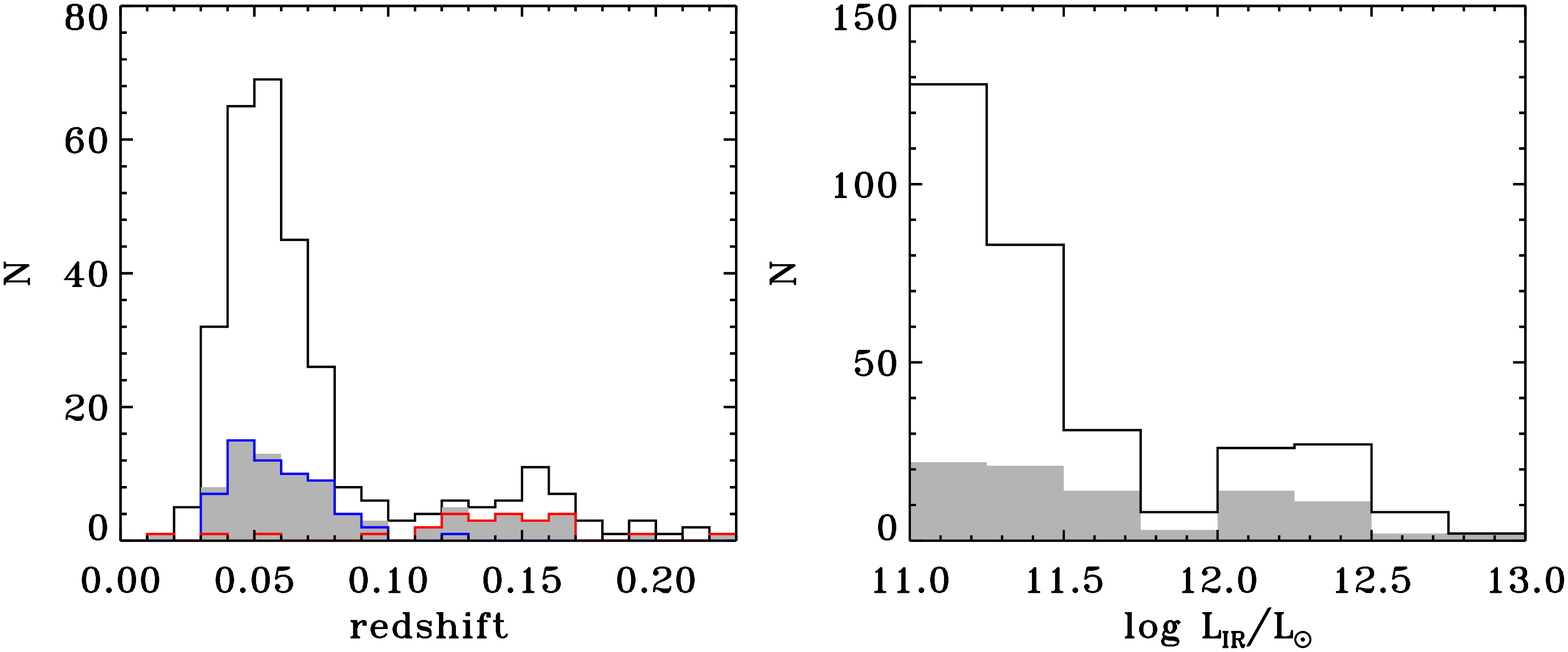}
\caption{Redshift and infrared luminosity distributions for (U)LIRGs.
The black histograms denote the 314 (U)LIRGs from cross-correlation
analysis, while the gray shaded histograms represent the 89
adv-merger (U)LIRGs. The blue and red histograms in the left panel represent
the redshift distributions of LIRGs and ULIRGs respectively.}
\label{zLirdis.eps}
\end{figure*}

Since spectral information is needed in our subsequent analysis, one ULIRG
without such information was excluded.  In addition, we removed 10 objects with
Seyfert 1 spectra from our sample, as the estimations for their
extinctions, stellar masses and star formation rates are unreliable.  Thus our
final adv-merger sample consists of 89 galaxies (60 LIRGs and 29 ULIRGs).
According to the commonly used
spectral classification BPT diagram proposed by Baldwin et al.  (1981) and
developed by Kauffmann et al. (2003a) and Kewley et al. (2001) based on
[\ion{N}{2}]~$\lambda 6584$/H$\alpha$ versus
[\ion{O}{3}]~$\lambda5007$/H$\beta$, we classified our sample galaxies into 25
star-forming galaxies, 42 composite galaxies and 20 narrow-line AGNs
(including Seyfert 2 and low-ionization narrow emission-line region galaxies).
For the remaining
two galaxies, the line fluxes have signal-to-noise ratios too low (S/N $<$ 5)
to be classified reliably. The black line and gray shaded histograms in Figure
\ref{zLirdis.eps} represent the redshift (left panel) and infrared luminosity (right panel)
distributions for the 314 (U)LIRGs before morphological classification and the
89 sample adv-merger (U)LIRGs, respectively. In the left panel of Figure \ref{zLirdis.eps}, we
also overplot the redshift distributions of LIRGs (blue line) and ULIRGs (red
line) separately. It can be seen that the lower redshift region ($z < 0.1$) is
dominated by LIRGs and the higher redshift region ($0.1 < z < 0.25$) is
occupied by ULIRGs.

\subsection{Control Samples}

For consistent estimates of physical parameters between our working sample and
the comparison samples, we do not use published relations in the literature.
Instead, we constructed the control samples by ourselves.  The control sample
used to investigate the SFR-$M_{\ast}$ relation for star-forming galaxies has
been extracted from the SDSS DR7 and satisfies $0.005<z<0.2$ and $14.5<r<17.77$. The
star-forming galaxies were selected according to the BPT diagram given by
Kauffmann et al. (2003a). The sample selection procedure is similar to that of
Brinchmann et al. (2004). This leaves us with 152137 galaxies.  For the
color-mass relation, the control sample was retrieved from
the Oh-Sarzi-Schawinski-Yi (OSSY) catalog (Oh et al. 2011)\footnote{
http://gem.yonsei.ac.kr/$\sim$ksoh/wordpress}, which is a spectroscopically selected
galaxy sample from the SDSS DR7 with redshifts z $<$ 0.2. It provides internal
extinction information (E(B-V)) from stellar continuum fits (EBV\_STAR).  For
comparison with Schawinski et al. (2014) in the color-mass relation studies, we
further restricted the redshift of the comparison sample to $0.02<z<0.05$ and
luminosity $M_{z,\rm Petro} < -19.5$ mag to obtain an approximately mass-limited sample as
described in Schawinski et al. (2014). The final control sample consists of 53604 galaxies.
 
\section{PARAMETER ESTIMATIONS}

In this paper, we mainly investigate the positions of adv-merger (U)LIRGs
in the SFR-stellar mass and the color-stellar mass diagrams. Hence we
need to carefully estimate the star formation rates, dust extinction corrected {\em
u-r}, {\em NUV-r} colors, as well as stellar masses for our sample
adv-mergers and the control sample galaxies.

The optical photometric data are all from SDSS DR7, while the near ultraviolet
({\em NUV}) magnitudes are from the {\em GALEX} satellite (Martin et al. 2005).
There are 79 sample adv-mergers and $\sim87\%$ control sample galaxies with
{\em NUV} photometric data available from the {\em GALEX}. We first performed
{\em k}-corrections to the optical and {\em NUV} magnitudes, followed by the
Galactic extinction and internal extinction corrections. The {\em
k}-corrections to the optical and {\em NUV} magnitudes were performed using the
New York University Value-Added Galaxy Catalog (NYU-VAGC; Blanton \& Roweis
2007) and by the IDL routine calc\_kcor.pro\footnote{http://kcor.sai.msu.ru/}
(Chilingarian et al. 2010; Chilingarian \& Zolotukhin 2012), respectively. The
foreground Galactic extinctions were corrected using dust maps from Schlegel et
al. (1998).  For the galaxies with SDSS spectroscopic information, the
EBV\_STAR values from the OSSY catalog (Oh et al. 2011) have been directly used
to correct for the internal dust extinctions. However, for 10 ULIRGs not in the
SDSS spectroscopic catalog and thus not in the OSSY catalog, we had to perform
internal extinction corrections ourselves, which is described as follows.
First, we calculated the color excess in the gas (EBV\_GAS) for the ULIRGs in
the SDSS spectroscopic survey using Balmer decrements H$\alpha$/H$\beta$,
assuming the case B recombination value of intrinsic H$\alpha$/H$\beta$ as 2.86
and 3.1 for star-forming galaxies and AGNs, respectively (Osterbrock \& Ferland
2006).  Second, we obtained the relation between EBV\_STAR and EBV\_GAS
for these galaxies: EBV\_STAR$=-0.03 + 0.61$EBV\_GAS.  Finally, we retrieved
the Balmer decrements for those ULIRGs without SDSS spectra from literature
(Veilleux et al. 1999; Darling et al.  2006) to calculate EBV\_GAS and derived
their EBV\_STAR using the EBV\_STAR and EBV\_GAS relation fitted above.  At
the last step, the dust extinction corrected {\em u-r} colors and {\em NUV}
magnitudes for all sample galaxies were obtained using the Calzetti et al.
(2000) extinction law and Cardelli et al. (1989) law, respectively, following
Schawinski et al.  (2014).  For our adv-merger sample, the median
internal extinction is 0.89 mag for the {\em u-r} color.

The stellar masses for most sample adv-mergers and all the control sample
galaxies were retrieved from the Max Planck Institute for Astrophysics-Johns
Hopkins University (MPA/JHU\footnote{http://www.mpa-garching.mpg.de/SDSS})
stellar mass catalog (Kauffmann et al. 2003b), which were estimated using
the SDSS five broad-band photometry.  Briefly, a median likelihood estimate of
the stellar mass was derived by comparing the five broad-band photometry with
the synthetic photometry of a large library of model star formation histories
using the Bayesian methodology. This library of model star formation histories
includes bursting and continuous star formation histories and covers a wide
range in metallicity.  In other words, this method has already taken proper
star formation histories into account during the modeling.  So for both
galaxies with bursting star formation histories and those with continuous star
formation histories, the stellar mass estimates should not suffer from
systematical impact from improper assumptions of their star formation histories.
For galaxies not included in the MPA/JHU stellar mass catalog, 10 out of 89
sample galaxies, we calculated the stellar masses following Bell et al. (2003),
utilizing the equation \begin{equation} \log (M_*/M_{\odot}) = -0.4(M_{\rm r,
AB} - 4.67)+[a_{\rm r} + b_{\rm r} (g-r)_{\rm AB} - 0.15], \end{equation} where
$M_*/M_{\odot}$ is the stellar mass in solar units, $M_{r,\rm AB}$ is the {\em
r}-band absolute magnitude in the AB magnitude system and $(g-r)_{\rm AB}$ is
the color in the rest-frame.  The coefficients $a_{r}$ and $b_{r}$ are taken as
$-0.306$ and 1.097, respectively, and the term $-0.15$ is adopted for the
Kroupa initial mass function (see Bell et al. 2003).  To check the consistency between these two
methods, we computed the stellar masses for the galaxies included in the
MPA/JHU stellar mass catalog using Equation (1).  The median
difference of the two approaches is 0.03 dex.  Therefore, the equation of Bell
et al. (2003) can be used statistically to measure the stellar masses for
galaxies not in the MPA/JHU catalog. Even though, we still use stellar masses
taken from MPA/JHU for most sample galaxies in this work, because for our dusty
starburst galaxies the stellar mass based on five broad-band
photometry and using the way described above should be more reliable for individual 
galaxy than that using only two band photometry by Bell et al. (2003).
 
Given that all our sample adv-mergers are infrared luminous and selected from the {\em IRAS}
database, the star formation rate can be derived from their total
infrared luminosity, following Kennicutt (1998, K98 hereafter)
\begin{equation}
\SFR({\rm IR})(\myear) = 4.5 \times 10^{-44} \LIR (\rm ergs\,s^{-1}).
\end{equation}

On the other hand, we can also calculate the star formation rate for star-forming and composite adv-mergers and 152137 star-forming control sample galaxies 
using their H$\alpha$ luminosities retrieved from the MPA/JHU catalog following K98
\begin{equation}
\SFR({\rm H}\alpha)(\myear) = 7.9 \times 10^{-42} L({\rm H}\alpha) (\rm ergs\,s^{-1}).
\end{equation}
The H$\alpha$ luminosities are aperture-corrected using the difference of the {\em r}-band Petrosian magnitude to
the fiber magnitude (Hopkins et al. 2003) as well as dereddened assuming the case B
recombination value of intrinsic H$\alpha$/H$\beta$ as 2.86.
Note that in the above formulae, the Salpeter (1955) initial mass function has been used and we converted them to the Kroupa one by dividing a factor 1.5.

\section{RESULTS}

\subsection{The SFR-Stellar Mass Relation}

It is generally accepted that star-forming galaxies lie on the MS
of the SFR-$M_*$ relation while starburst galaxies are located above
the MS (Rodighiero et al. 2011; Hung et al. 2013). Below we investigate where
our sample adv-mergers are located in the same diagram to obtain clues on their
evolution.

Figure \ref{msLir.eps} shows the SFR(IR)-$M_*$ relation for 89 sample adv-mergers.  The
ULIRGs and LIRGs are shown as red and blue filled circles, respectively.  The
black solid and dashed lines in Figure \ref{msLir.eps} represents the best fit MS relation
and 1$\sigma$ scatter after 3$\sigma$ clipping of outliers, respectively, based
on 152137 star-forming galaxies selected from SDSS DR7 (see Section 2.2) as
\begin{equation} 
\log {\rm SFR}({\rm H}\alpha) = (1.02\pm0.001)\log
(M_*/M_{\odot}) - (10.01\pm0.014).  
\end{equation} 
The $1\, \sigma$ dispersion of our fitting MS relation is 0.3 dex after the
typical SFR estimation scatter $\sim$ 0.2 dex is removed, which is consistent
with the scatters derived in the literature (e.g. Noeske et al. 2007;
Rodighiero et al.  2011; Whitaker et al. 2012; Sargent et al. 2012).
Note that for the normal star-forming galaxies used to fit the MS relation, their SFRs are calculated with their H$\alpha$ luminosities, which are consistent with those estimated from their infrared luminosities (Kennicutt et al. 2009; Lee et al. 2013).

It is clear from Figure \ref{msLir.eps} that almost all adv-merger (U)LIRGs are above MS line
and a large fraction ($74\%\pm5\%$) of them are localized above the $1\,
\sigma$ line of MS relation.  In particular, the ULIRGs are far above the MS
line as outliers, consistent with  previous findings (e.g. Elbaz et al. 2007;
da Cunha et al. 2010). The fraction of LIRGs above MS relation is also high
($62\%\pm6\%$). Given that the definition of starburst galaxies based on deviations 
from the MS relation varies in the literature (Rodighiero et al. 2011; Hung et
al. 2013), in this work we define starburst galaxies as those above the $1\,
\sigma$ line of MS relation. Then more than two-thirds ($74\%$) of our sample
adv-merger (U)LIRGs are in the starburst phase. In practice, this definition of
starburst is also consistent with starburst definitions based on other
starburst indicators, such as high star formation rates and short molecular gas
exhausting timescales (Knapen \& James 2009; see Section 5.3).

\subsection{The Color-Stellar Mass Relation}

\subsubsection{The u-r Color-Mass Relation}

Since our goal is to investigate the role of gas-rich major
mergers on galaxy evolution from the blue cloud to the red sequence of quiescent
galaxies in the color-mass diagram,
accurate dust extinction correction is very important.
Schawinski et al. (2014) found
that the galaxies in the blue cloud are bluer in the dust extinction corrected
color-mass diagram compared with the uncorrected one, leading to a more prominent
blue cloud and red sequence. For our adv-merger (U)LIRGs sample, dust extinction is
even more serious: the median internal extinction is 0.89
mag on the {\em u-r} color as derived in Section 3. Therefore, we performed
both foreground Galactic and galaxy internal extinction corrections carefully.

\begin{figure*}
\plotone{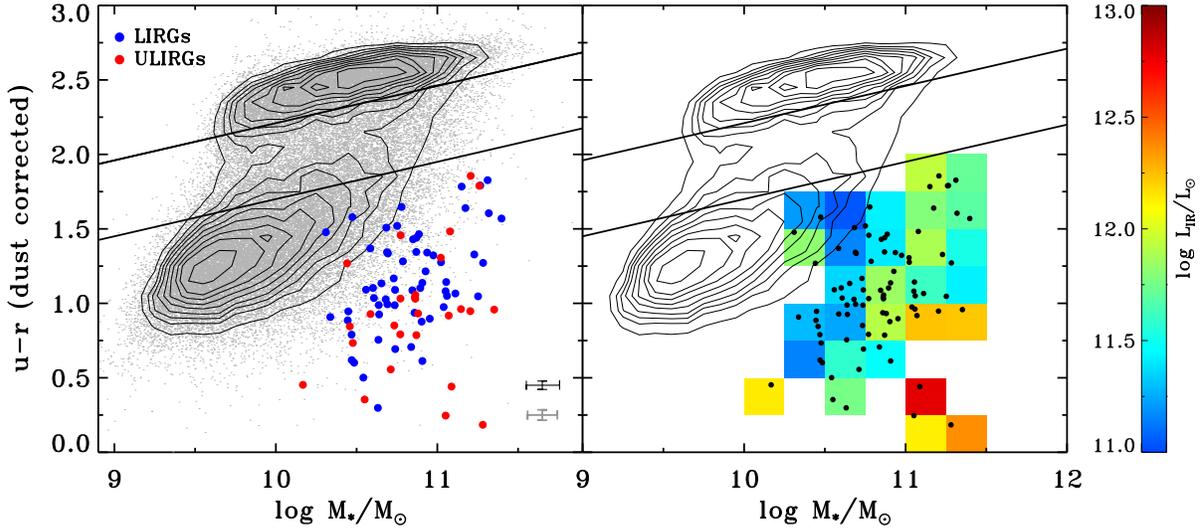}
\caption{Left panel: dust-corrected {\em u-r} color-mass diagram for sample
adv-mergers. Blue and red filled circles represent LIRGs and ULIRGs,
respectively. The contours show 9 equally spaced levels between 10\% and 90\%
number densities of galaxies drawn from SDSS DR7 (gray dots).
The black and gray error bars show the median uncertainties for the sample adv-mergers
and the control sample galaxies, respectively. The black solid
lines are taken form Schawinski et al. (2014) as the separation between the
blue cloud, green valley and red sequence and have been shifted downward by
0.07 mag to account for foreground Galactic extinction correction.
Right panel: same as the left panel but with the $\LIR$ distribution of these
adv-mergers overlaid. The color scale of $\LIR$ is shown to the right.}
\label{cmdadv_ur.eps}
\end{figure*}

Figure \ref{cmdadv_ur.eps} illustrates the foreground Galactic and internal dust extinction
corrected {\em u-r} color-mass diagram. In the left panel of Figure \ref{cmdadv_ur.eps}, the gray
dots represent 53604 comparison sample galaxies (see Section 2.2), and the
overlaid contours show the galaxy number densities for these galaxies. The
lowest density contour represents the 90\% number density. The control sample
galaxies show an obvious bimodal distribution as the blue cloud and the red
sequence.  The area in between is the green valley. The two black lines in
Figure \ref{cmdadv_ur.eps} denote its boundaries, which are similar to the definition of
Schawinski et al. (2014)
\begin{equation}
u-r = -0.24 + 0.25 \log (M_*/M_{\odot}),
\end{equation}
\begin{equation}
u-r = -0.75 + 0.25 \log (M_*/M_{\odot}).
\end{equation}

Note that the black lines shown in Figure \ref{cmdadv_ur.eps} are 0.07 mag bluer than the above
definition to account for the Galactic extinction corrections we performed.
The median Galactic extinction for the comparison
sample is $\sim$ 0.07 mag. Since the boundary definition of the green
valley is somewhat arbitrary, the adjustment we took is more suitable for
the galaxy distribution in the color-mass diagram shown in Figure \ref{cmdadv_ur.eps} than those
given by Equations (5) and (6).

The red and blue filled circles in the left panel of Figure \ref{cmdadv_ur.eps} 
represent adv-merger ULIRGs and LIRGs, respectively.  
It is clear from the left panel of Figure \ref{cmdadv_ur.eps} that
almost all our sample galaxies are indeed localized to the right of blue cloud galaxies
in the color-mass diagram and $95\%\pm2\%$ sample galaxies are outside the lowest density contour due to 
their larger stellar masses and bluer colors
(simply refer to ``at the bottom right corner of the color-mass diagram'', hereafter). 
Therefore, there do exist a distinct population of galaxies that belong to neither the blue cloud, red sequence,
nor the green valley. Actually, Chen et al. (2010) have already reported this population in
the {\em g-r} color-mass diagram for 52 local galaxies selected from {\em IRAS} 1 Jy ULIRG sample,
but only $\sim 46\%$ of their ULIRGs are located outside the $90\%$ number density contour.
From the analysis of a larger sample with 82 local ULIRGs from cross-matching the {\em AKARI} all-sky survey (Murakami et al. 2007) 
with SDSS DR10 (Ahn et al. 2014) and the Two-Degree Field Galaxy Redshift Survey (2dFGRS; Colless et al. 2001), Kilerci Eser et al. (2014)
obtained similar results to that of Chen et al. (2010).
We note that both studies have not applied internal dust extinction corrections,
which is the main cause for the difference in the fractions of galaxies outside the $90\%$ density contour between this work
and those of Chen et al. (2010) and Kilerci Eser et al. (2014). 

We also examine the $\LIR$ distribution of our adv-merger (U)LIRGs in the color-mass
diagram in the right panel of Figure \ref{cmdadv_ur.eps} by different colors. It is clear from
the right panel of Figure \ref{cmdadv_ur.eps} that the galaxies tend to be located further away
from the blue cloud area as the $\LIR$ increases. 

Statistical studies based on large samples have not noticed this population,
implying that the fraction of such galaxies is small in the local universe.  It
should be interesting to have a rough order-of-magnitude estimate for the
fraction.  To derive this fraction, we first calculated the number density of
(U)LIRGs based on the analytical infrared luminosity function (LF) provided by
Goto et al. (2011). The integrated number density  over the luminosity range of
$10^{11} L_\odot$ to $10^{12.5} L_\odot$ is $2.1 \times 10^{-5} \Mpc^{-3}$,
after converting the cosmology adopted by Goto et al. (2011) to the one used in
this work. We note that the infrared luminosity function constructed by Goto et
al. (2011) only probes galaxies with luminosities from $10^{9.8}L_\odot$ to
$10^{12.5} L_\odot$ while many blue cloud galaxies may have infrared
luminosities fainter than $10^{9.8}L_\odot$. So an integrated number density
over this infrared luminosity range would under-estimate the total number
density of blue cloud galaxies. To remedy this, we resort to an optical
luminosity function to estimate the number density of blue cloud galaxies: we
used the {\em r}-band luminosity function for late-type galaxies by Nakamura et
al. (2003) since blue cloud galaxies are mostly late-type (Schawinski et al.
2014).  The total number density in the luminosity range of
$-23.3<M_{r}<-18.8$, which brackets the blue cloud galaxies, is $8.2 \times
10^{-3} \Mpc^{-3}$.  Consequently, the fraction of (U)LIRGs to late-type
galaxies is about 0.26\%.  Note that although ULIRGs are almost exclusively
interacting or merging systems, about half of the LIRGs are spirals (Wang et
al. 2006). Therefore the real fraction of infrared luminous adv-mergers should
be smaller by a factor of a few, i.e., less than 0.1\%. 
Considering that late-type galaxies may include galaxies not in the blue cloud, this fraction may 
be higher than 0.1\%.  But it will not change in the order-of-magnitude.  
Such a tiny fraction suggests that blue, massive dusty galaxies at the bottom right
corner of the color-mass diagram are very rare in the local universe.

\begin{figure*}
\plotone{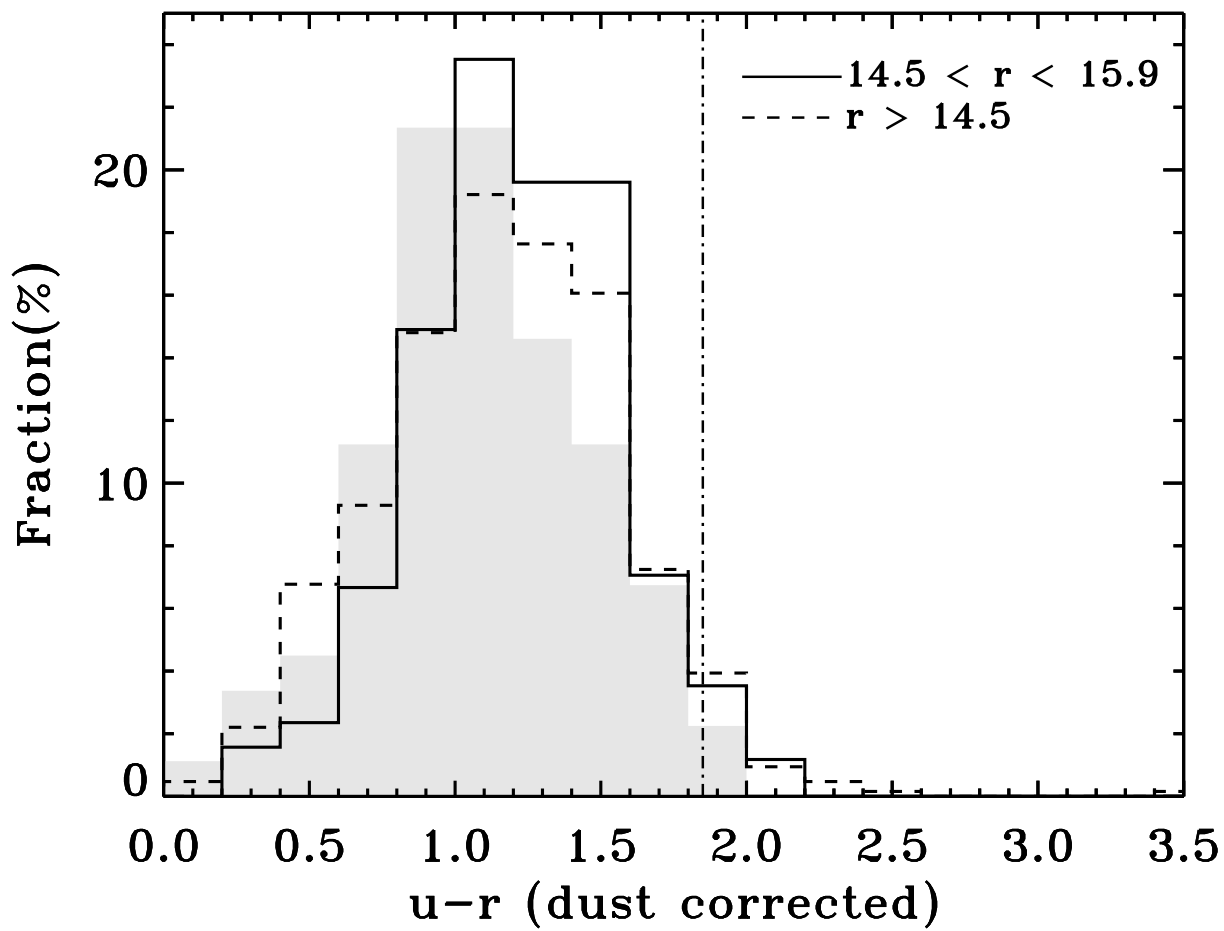}
\caption{Comparisons of the {\em u-r} color distributions between the (U)LIRGs selected from
the cross-correlation analysis with $14.5<r<15.9$ (solid line) and those with $r>14.5$
(dashed line). The gray shaded histogram represents the 89 sample adv-merger
(U)LIRGs. The black vertical dot-dashed line denotes the color of the
lower boundary of the green valley at the median stellar mass of our sample
galaxies.}
\label{urdistri_raw2.eps}
\end{figure*}

Considering that we have put a constraint on the magnitude range of
$14.5<r<15.9$ when we selected our (U)LIRGs sample by cross-correlation
analysis, we should examine possible selection effects by such a magnitude cut.
Figure \ref{urdistri_raw2.eps} shows the comparison of {\em u-r} color distributions for 259 (U)LIRGs within the
magnitude range of $14.5<r<15.9$ (solid line) and 636 (U)LIRGs with $r>14.5$
(dashed line).  The black vertical dot-dashed line denotes the {\em
u-r} color for the lower boundary of the green valley at the median stellar
mass of our sample galaxies. We can see from Figure \ref{urdistri_raw2.eps} that almost all (U)LIRGs
are blue, and the galaxies in magnitude ranges of $14.5<r<15.9$ and those with $r>14.5$ cover
similar ranges in {\em u-r} colors, indicating that the constraint on magnitude
range of $14.5<r<15.9$ does not preferentially select blue galaxies.
In addition, the gray shaded histogram in Figure \ref{urdistri_raw2.eps} represents the {\em u-r}
color distribution for 89 sample (U)LIRGs, from which we can see that the
adv-merger (U)LIRGs are slightly bluer than the whole (U)LIRGs sample. 

\subsubsection{The NUV-r Color-Mass Relation}

\begin{figure*}
\plotone{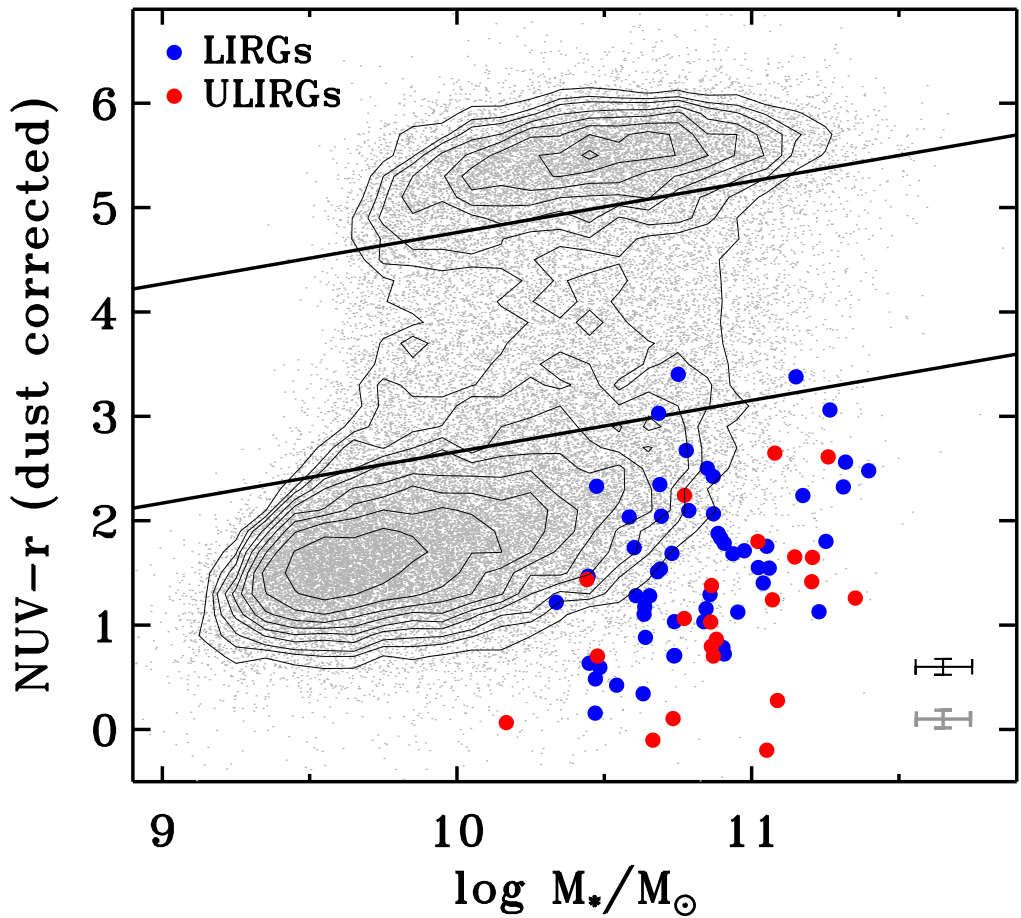}
\caption{Dust-corrected {\em NUV-r} color-mass diagram for sample adv-mergers
with {\em NUV} photometry from {\em GALEX}.
Blue and red filled circles represent LIRGs and ULIRGs, respectively.
The contours show 9 equally spaced levels between 10\% and 90\% number densities of 46407 galaxies
drawn from SDSS DR7 with {\em NUV} photometry (gray dots). The black and gray
error bars show the median uncertainties for the sample adv-mergers and the
control sample galaxies, respectively. The black solid lines indicate
the boundaries of the green valley according to Equations (7) and (8).}
\label{cmdadv_NUVr.eps}
\end{figure*}

It is well known that the rest-frame optical colors are sensitive to star
formation on timescales of $10^8-10^9$ yr (K98; Hopkins et al. 2003;
Moustakas et al. 2006) while the ultraviolet emission traces the star-forming process on
shorter timescales of $10^7-10^8$ yr, thus the {\em NUV-r} color is more closely
related to the current star formation process (K98).
Actually, Cortese (2012) found that optically red massive spirals are on the
UV-optical blue cloud and concluded that it is difficult to
distinguish active star-forming galaxies from truly passive ones using
optical colors alone. Figure \ref{cmdadv_NUVr.eps} shows the dust-corrected {\em NUV-r} color versus stellar mass
relation for 79 sample galaxies and 46407 comparison sample galaxies. 
We use a similar method to that used for the {\em u-r} color-mass diagram to define
the boundaries of the green valley in the {\em NUV-r} color-mass diagram as
follows:
\begin{equation}
NUV-r = -0.159 + 0.492 \log (M_*/M_{\odot}),
\end{equation}
\begin{equation}
NUV-r = -2.259 + 0.492 \log (M_*/M_{\odot}).
\end{equation}

In comparison with Figure \ref{cmdadv_ur.eps}, the separation of the blue cloud and the red
sequence is more prominent and the green valley covers a wider range in {\em
NUV-r}, which is consistent with previous studies (Martin et al.
2007; Salim et al. 2007; Schiminovich et al. 2007; Wyder et al.  2007). 
Upon a closer examination of the relative positions of our sample galaxies to the blue cloud, 
we find a difference from those shown in Figure \ref{cmdadv_ur.eps}. 
Although most ($81\%\pm4\%$) of our sample galaxies are still localized at the bottom right
of the {\em NUV-r} color-mass diagram and do not reside in the blue cloud, a
few sample galaxies show up in the green valley in the {\em NUV-r} color-mass
diagram. 

\section{DISCUSSION}

\subsection{Stellar Mass Estimates for (U)LIRGs}
 
The stellar mass is a fundamental quantity and is a key parameter in this work.
The stellar masses of our sample galaxies are estimated via SED modeling.
Due to the complex star formation histories of young stellar populations in
(U)LIRGs (e.g. Rodr{\'{\i}}guez Zaur{\'{\i}}n et al.  2007, 2008),  the stellar
masses of (U)LIRGs estimated via SED modeling may have large uncertainties and systematic biases.

\begin{figure*}
\plotone{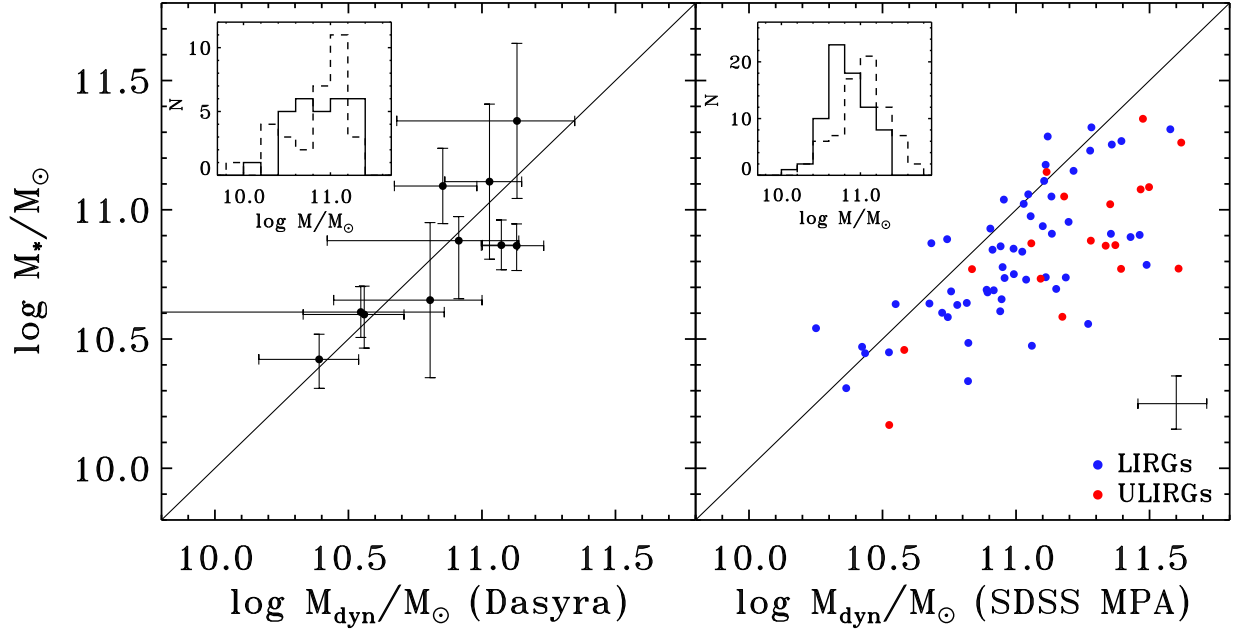}
\caption{Comparisons between the stellar masses ($\rm M_*$) and the dynamical
masses ($\rm M_{dyn}$).  Left panel: $\rm M_*$ vs.  $\rm M_{dyn}$ for 10
overlapping ULIRGs between Dasyra et al. (2006a, 2006b) and our sample. The solid line
represents the line of equality.  The error bars show the uncertainties for
each individual points. The inset histograms show the distributions of $\rm
M_*$ for our 29 sample ULIRGs (solid line) and the $\rm M_{dyn}$ for 31 ULIRGs
with single nucleus from Dasyra et al.  (2006b) (dashed line). The median logarithmic values
of these two distributions are 10.86 and 10.91 respectively. Right panel: $\rm
M_*$ vs. $\rm M_{dyn}$ for 74 sample adv-merger (U)LIRGs with velocity
dispersions from the MPA/JHU catalog.  The blue and red dots represent LIRGs
and ULIRGs, respectively. The solid line represents the line of equality. The
error bars show the median uncertainties.  The inset histograms show the
distributions of $\rm M_*$ (solid line) and $\rm M_{dyn}$ (dashed line) for the
74 sample adv-merger (U)LIRGs. The median values of these two distributions for
$\rm \log M_*/M_\odot$ and $\rm \log M_{dyn}/M_\odot$ are 10.84 and 11.06, respectively.}
\label{mdyn.eps}
\end{figure*}

We compare the stellar masses estimated in this work with the dynamical masses
of ULIRGs by Dasyra et al. (2006a, 2006b). Dasyra et al. (2006a, 2006b) provided the largest
ULIRG sample (54) with dynamical mass measurements, based on high resolution,
long-slit {\em H-} and {\em K-} band spectroscopy. Their sample consists of 31
single nucleus and 23 double nuclei ULIRGs. We retrieve the dynamical masses
from Dasyra et al. (2006a, 2006b) for 10 ULIRGs that overlap with our sample.  The left
panel of Figure \ref{mdyn.eps} compares the stellar masses of the 10 ULIRGs measured by their SEDs
with their dynamical masses derived from CO ro-vibrational line measurements.
Although the sample size is small, it is obvious that the stellar
masses and the dynamical masses are consistent with each other and the largest difference is
only 0.3 dex. In the inset, we compare the
distributions of stellar masses for our sample ULIRGs and the dynamical masses
of 31 single nucleus ULIRGs retrieved from Dasyra et al. (2006b).
These two distributions are quite similar with nearly identical median logarithmic values
(10.86 and 10.91, respectively), and the K-S test gives a
probability of 0.67, indicating that the differences between these two
distributions are not statistically significant. Therefore, it appears that 
the stellar masses measured via SED modeling are reliable for local ULIRGs.

To extend the size of sample (U)LIRGs with dynamical masses ($\rm M_{dyn}$), we
also estimate the dynamical masses approximately by $\rm M_{dyn}$ = $5r_{50}\, \sigma ^2/\rm
G$ (e.g., Bernardi et al. 2010), where the velocity dispersion ($\sigma$) and
the S{\'e}rsic half-light radius $r_{50}$ are retrieved from the MPA/JHU catalog
and NYU-VAGC (Blanton et al. 2005), respectively.  After excluding the objects
with low signal-to-noise ratios (S/N $<$ 10) and with velocity dispersions less
than 70 ${\rm km/s}$ that is the resolution limit of SDSS spectra, there are 74
sample galaxies (18 ULIRGs and 56 LIRGs) with velocity dispersions and $r_{50}$
available from the  MPA/JHU catalog.  We compare the stellar masses with the
dynamical masses for these 74 sample adv-merger (U)LIRGs in the right panel of
Figure \ref{mdyn.eps}. As can be seen, the stellar masses are systematically smaller
than the dynamical masses. In the inset, we compare the distributions of
stellar masses and the dynamical masses for the 74 sample (U)LIRGs.
The median logarithmic values of the two distributions are
10.84 and 11.06, respectively, indicating that the stellar masses
are $\sim 0.2$ dex smaller than the dynamical masses for sample (U)LIRGs based
on the SDSS database.

Given that the above two comparisons based on different datasets yield slightly
different results, we compare the data for the 8 overlapping sample (U)LIRGs with
dynamical properties measured by Dasyra et al. (2006a, 2006b) and SDSS.  We find that
the effective radii from Dasyra et al. (2006a, 2006b) are smaller than those of SDSS.
In fact, Veilleux et al. (2002) have already found that the effective radii
measured in the {\em K-}band are smaller than those in the {\em R-}band for {\em IRAS} 1 Jy
ULIRGs samples, due to the extremely dusty environment for the ULIRGs.  
In contrast, the velocity dispersion measurements have no significant systematic
differences from different wavebands (Dasyra et al. 2006b).
Therefore, the dynamical masses for 74 sample (U)LIRGs measured by SDSS may be
at some level overestimated.

From the comparisons of stellar masses with the dynamical masses from different 
datasets, we conclude that the stellar masses measured via SED modeling are
not over-estimated for our sample (U)LIRGs. Therefore, the adv-merger (U)LIRGs are
located in the bottom right corner of the color-mass diagrams.

\subsection{Star Formation Rate Estimates}

\begin{figure*}
\plotone{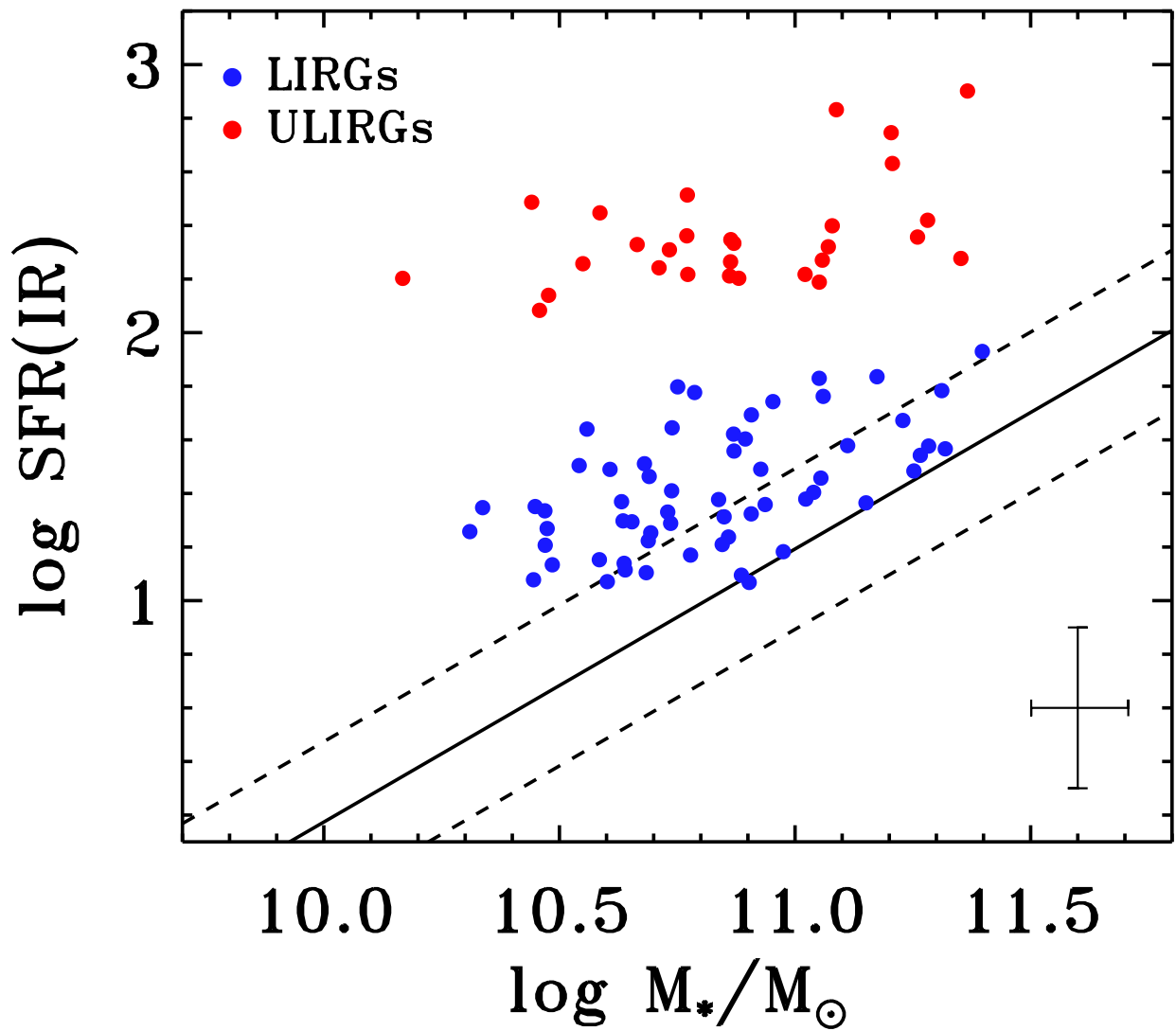}
\caption{SFR(IR) vs. stellar mass relation for our sample adv-mergers.
Blue and red filled circles represent LIRGs and ULIRGs, respectively.
The error bars show the median uncertainties. The black
solid line is the local main sequence (MS) relation based on 152137
star-forming galaxies of SDSS DR7 and SFR(H$\alpha$) are used. The dashed
lines show the 1 $\sigma$ dispersions (0.3 dex) from the MS fitting.}
\label{msLir.eps}
\end{figure*}

When we investigate the locus of our sample adv-merging (U)LIRGs in the
SFR-$M_*$ diagram, the SFRs were estimated using their infrared luminosities.
However, Hayward et al. (2014) argued based on their hydrodynamical simulations
that for strong starburst galaxies at the post-starburst phase, the
instantaneous SFR calculated from the infrared luminosity can be severely
over-estimated. This is because the contribution from OB stars to dust heating
decreases (the lifetime of OB stars is only $10^{6}$ to $10^{7}$ years) and
that from older stellar populations becomes non-negligible when the SFR starts
to decrease (e.g., Kennicutt et al. 2009). Therefore the infrared luminosity is
not a good indicator of the instantaneous SFR for galaxies at the
post-starburst phase, although it is the best SFR tracer for extremely dusty starbursts.
On the other hand, the most commonly used H$\alpha$ luminosity can
serve as a good SFR indicator for normal star-forming galaxies, for which dust
attenuation can be corrected properly. However, for dusty starburst galaxies,
massive starbursts tend to take place in obscured circumnuclear regions (K98) where the
optical depths are generally an order of magnitude higher than the upper limit of
the reddening correction for H$\alpha$ ($\sim$2.5 mag) (Lonsdale et al. 2006;
Moustakas et al. 2006). It is impossible to correct internal dust extinctions
accurately for such high optical depths with the Balmer decrements. In such cases,
the H$\alpha$ luminosity will under-estimate the real SFRs for dusty starburst
galaxies. Given that our galaxy sample consists of both galaxies with transiting post-starburst features (see Section 5.3) and
dusty starbursts, it is very likely that the real SFRs of our sample galaxies are in
between the SFR(IR) and SFR(H$\alpha$). Thus, it is worth investigating
the positions of our galaxies on the SFR(H$\alpha$)-$M_*$ diagram to see how much
will be changed compared to the SFR(IR)-$M_*$ diagram shown in Figure \ref{msLir.eps}.

\begin{figure*}
\plotone{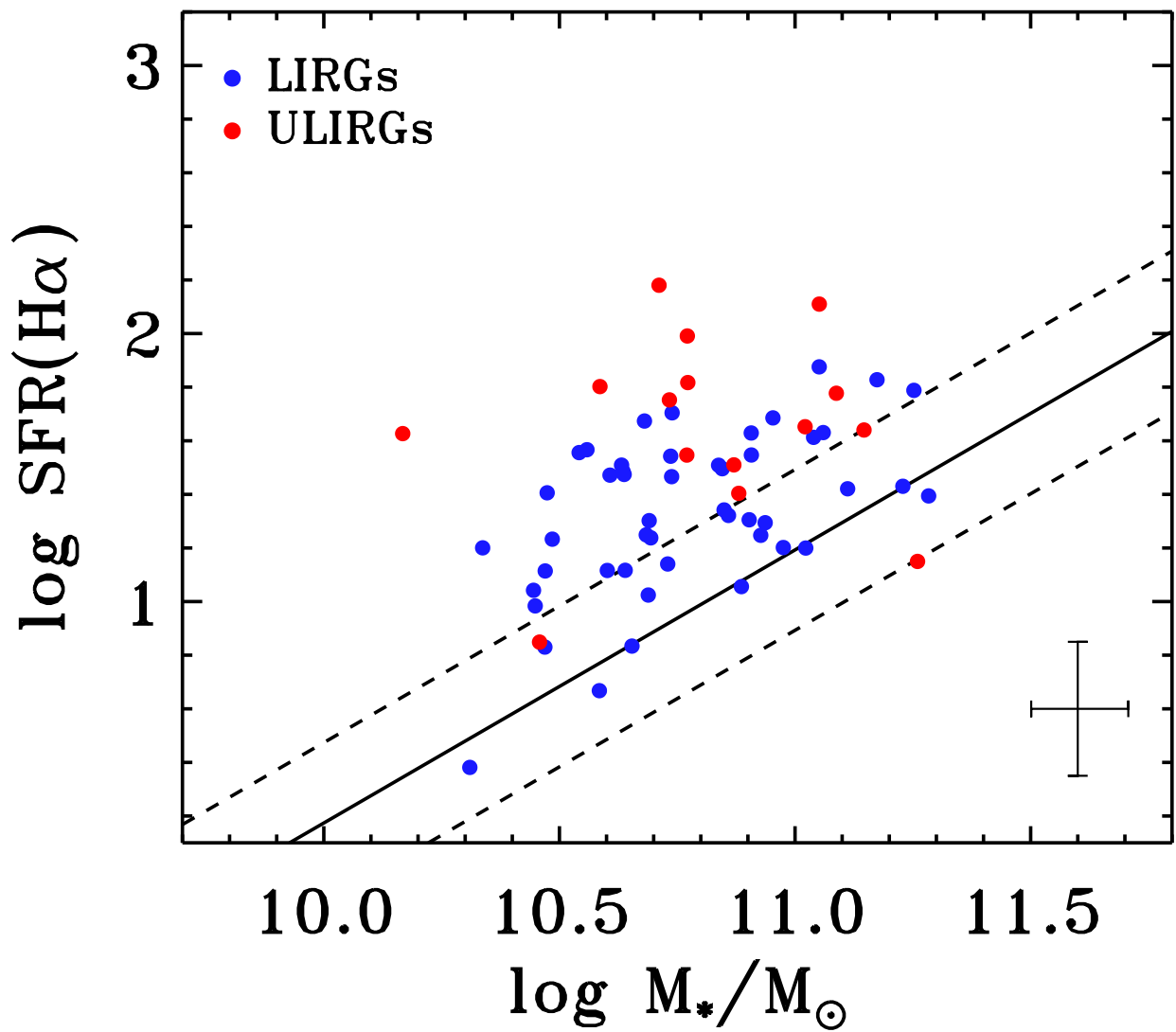}
\caption{SFR(H$\alpha$) vs. stellar mass relation for star-forming and
composite sample adv-mergers. The main
sequence (MS) line and symbols are the same as in Figure \ref{msLir.eps}.
The median error bars are shown in the bottom right.}
\label{msHa.eps}
\end{figure*}

We plot the SFR(H$\alpha$) versus stellar mass in Figure \ref{msHa.eps}, which is the same
as Figure \ref{msLir.eps} except that the H$\alpha$ luminosity is used as the SFR indicator.
It is clear from Figure \ref{msHa.eps} that the positions of our sample (U)LIRGs in the
SFR(H$\alpha$)-$M_*$ are quite different from those in Figure \ref{msLir.eps}. In particular,
ULIRGs are located just above the MS relation and mixed with LIRGs as shown in
Figure \ref{msHa.eps}, rather than being apparent outliers as in Figure \ref{msLir.eps}.  Even so, most
of our dusty adv-merging (U)LIRGs are still located above the MS line in both Figures
\ref{msLir.eps} and \ref{msHa.eps}, for which SFRs are estimated from $\LIR$ and L(H$\alpha$),
respectively. Therefore the use of different SFR indicators
does not influence our main conclusions reached in Section 4.1.

\subsection{How do Adv-merger (U)LIRGs Evolve on the Color-Mass Diagrams?}

The morphologies and locus of our sample adv-merger (U)LIRGs on the
SFR(IR)-$M_*$ relation as shown in Figure \ref{msLir.eps} suggest that most of
them are experiencing massive starbursts as triggered by gas-rich major
mergers. These galaxies are blue and massive with stellar mass
$M_{\ast}>3\times 10^{10} M_{\odot}$; they are located to the right of the blue
cloud galaxies on the optical and NUV color-mass diagrams, implying that they
belong to a distinct population being neither in the blue cloud, red sequence
nor the green valley. In spite of the limited data, it  is still interesting
to investigate how these adv-merger (U)LIRGs may evolve on the color-mass
diagrams. To explore this issue, two questions naturally arise: 
What is the timescale for adv-merger (U)LIRGs exhausting their molecular gas, 
and then evolving to post-starbursts and finally being quenched?  
Are there transiting objects from starbursts to post-starbursts among our sample galaxies?

The starburst duration of adv-merger (U)LIRGs can be simply quantified by the
gas consumption timescale, i.e., the available gas mass divided by the
current SFR.  As already pointed out by Daddi et al. (2010) and Genzel et  al.
(2010) based on CO observations for both low and high redshift galaxies, the
molecular gas consumption is bimodal. Disk galaxies consume their gas reservoir
slowly, while in starburst galaxies the gas is rapidly exhausted. The fast mode occurs 
because during a gas-rich major merger gas falls
toward the central region and starbursts take place at central kpc rapidly due to
high dense gas fractions (Downes \&  Solomon 1998; Barnes \&  Hernquist 1996).
Given that only 8 of our sample adv-merger (U)LIRGs have existing CO data, we
examine the molecular gas depletion timescales mainly based on the adv-merger
(U)LIRGs sample by Liu et al. (2015). They collected 66 (U)LIRGs with CO data
from the literature and derived their molecular gas masses, out of which 35
(U)LIRGs are adv-merging galaxies. The molecular gas masses of these 35 adv-merging
(U)LIRGs plus 6 of our sample (U)LIRGs\footnote{Out of the 8 of our adv-merger
(U)LIRGs with existing CO observations, 2 are already included in Liu et al. (2015).}
are in the range of about $8\times 10^8$ $M_{\odot}$ to
$2\times 10^{10}$ $M_{\odot}$ where we use the conversion factor  (0.8) from
the CO luminosity to the molecular gas  mass for local (U)LIRGs 
(Solomon \&  Vanden  Bout 2005). We find the timescale of gas consumption in these
adv-merger (U)LIRGs ranges from  $3\times 10^{7}$  to  $3\times
10^{8}$  years, consistent with the rapid star formation mode for starburst
galaxies seen by Daddi et al.  (2010) and Genzel et al. (2010).

Furthermore, to investigate the link between starbursts and post-starbursts, and
to distinguish them from slow quenching normal galaxies, Yesuf et al. (2014)
invented two plausible criteria for identifying transiting post-starbursts from
a EW(H$\alpha$) and EW(H$\delta_A$) restricted\footnote{3{\AA} $<$
EW(H$\alpha$) $<175${\AA} and EW(H$\delta$$_A$) $>4${\AA}} sample: The first is
to select the ``fading post-starbursts'' based on a 3D parameter space
constructed by dust-corrected {\em NUV-g} color, EW(H$\delta$$_A$) as well as the
4000{\rm \AA} break, D$_n(4000)$; the second is to identify the ``obscured post-starbursts''
using the {\em GALEX} and {\em Wide-field Infrared Survey Explore} ({\em WISE};
Wright et al. 2010) photometry, which is likely the bridge between starbursts
and ``fading post-starbursts''.  Specifically, ``fading post-starbursts'' are
galaxies with more than 2$\sigma$ deviations from normal galaxy locus in the 3D
parameter space constructed by EW(H$\delta$$_A$), 
D$_n(4000)$ and {\em NUV-g} color, while the ``obscured post-starbursts'' are
defined as galaxies with $f_{12 \mu m}/f_{0.2 \mu m} > 200$ and $f_{4.6 \mu
m}/f_{3.4 \mu m} > 0.85$, where $f_{12 \mu m}/f_{0.2 \mu m}$ is the flux
density ratio between {\em WISE} 12$\mu$m and {\em GALEX NUV} 0.2$\mu$m, and
$f_{4.6 \mu m}$/$f_{3.4 \mu m}$ is the flux ratio between {\em WISE} 4.6$\mu$m
and 3.4$\mu$m.

According to the criteria\footnote{Instead of using the {\em NUV-g} color, we
use our dust-corrected {\em NUV-r} color measurements for the selection of
``fading post-starbursts''  for convenience.} proposed by Yesuf et al. (2014), 
there is no ``fading post-starbursts'' found in our sample.  However, based on the {\em GALEX}
and{\em WISE} photometric data, we identified 12 sample galaxies satisfying the
criteria of the ``obscured post-starbursts'' (see Table 1). From a rough comparison of their SEDs
with the stellar population synthesis models,  Yesuf et al. (2014) found that 
the ``obscured post-starbursts'' have age of about  400-500 Myr since the last
starburst epoch, and are younger than the ``fading post-starbursts''.
Therefore, although most of our sample galaxies are still experiencing
massive starbursts, 12 ($\sim 14\%$) of them are in the initial stage of
transiting post-starbursts, which is consistent with the theoretical prediction
that post-starburst galaxies are the end products of gas-rich major mergers
after their gas has been exhausted (Hopkins et al. 2006, 2008).

The short molecular gas depletion timescale along with an  appreciable fraction
of obscured transiting post-starbursts of our sample favor the scenario that
the adv-merger (U)LIRGs follow the fast evolutionary track proposed  by
several groups (e.g.  Muzzin et al. 2013a;  Barro et al. 2013, 2014a, 2014b;
Marchesini et al. 2014; Schawinski et  al. 2014; Belli et al. 2015 and Wellons
et  al. 2015): the massive gas-rich disk galaxies experience violent merging
processes and evolve from the blue cloud  galaxies to massive blue and dusty
galaxies (from  left to right on the color-mass diagram). After the molecular
gas has been rapidly exhausted and their star  formation ceases, these
adv-mergers become massive red-sequence galaxies through the transiting
post-starburst phase within $\sim$ 1 Gyr.

\section{SUMMARY}

With the aim of exploring the evolutionary pathway from the blue cloud of star-forming galaxies 
to the red sequence of quiescent galaxies, we use a local adv-merger (U)LIRGs sample 
and apply careful dust extinction corrections to investigate their positions 
in the SFR-$M_{\ast}$, {\em u-r} and {\em NUV-r} color-mass diagrams.
Our sample consists of 89 luminous infrared and ultraluminous infrared galaxies at the late merger stage,
which has been obtained from cross-correlating the {\em IRAS} PSCz and 1 Jy ULIRGs
samples with the SDSS DR7 database. Our main results are summarized as
follows.

\begin{enumerate}
\item We find that almost all adv-merger (U)LIRGs are located above the local
star-forming galaxy main sequence (MS) in the SFR-$M_{\ast}$ diagram.
Specifically, the ULIRGs are far above the MS line as outliers and $62\%\pm 6\%$ LIRGs
are above the $1\, \sigma$ line of the MS, indicating that a majority (two-thirds) of these
gas-rich major merger galaxies at the late merging stage are at the starburst stage 
with high star formation rates.

\item We also find that almost all our sample adv-merger (U)LIRGs
are as blue as the blue cloud galaxies after carefully removing the Galactic and internal
dust extinctions. They are also massive and thus localized to the right of the
blue cloud galaxies on
the {\em u-r} and {\em NUV-r} color-mass diagrams with $95\%\pm 2\%$ and $81\%\pm 4\%$ sample galaxies 
outside the $90\%$ density contour of blue cloud galaxies, respectively.
Therefore, the adv-merger (U)LIRGs are a distinct population in the color-mass diagram 
that differ from the blue cloud, green valley, and red sequence galaxies. 
However, these galaxies are rare in the local universe and their number density is only $\sim 0.1\%$ 
of that of the local blue cloud galaxies. In contrast, the fraction of similar
objects at high redshift is much larger (7\%) (see Straatman et al. 2015).

\end{enumerate}

The molecular gas depletion timescale for adv-merger
(U)LIRGs is $3\times 10^{7}$ to $3\times 10^{8}$ years, which is much shorter
than that for disk galaxies in the blue cloud.  We also identified 12 ($\sim 14\%$)
obscured transiting post-starbursts based on the {\em GALEX} and {\em WISE}
photometry following Yesuf et al. (2014), implying that more than ten percent
of adv-merger (U)LIRGs are at the initial stage of transiting post-starbursts.
Therefore, our results  favor the fast evolutionary track proposed by several
groups (e.g. Marchesini et al. 2009, 2014; Brammer et al. 2011; Muzzin et al. 2013a;
Barro et al. 2013, 2014a, 2014b; Schawinski et al. 2014; Belli et al. 2015; Mancini et
al. 2015) that massive disk galaxies experience violent merger processes and
evolve from the blue cloud first to the massive blue and dusty galaxies, and then rapidly
to the red sequence through the green valley and eventually form massive quiescent
galaxies. Given that the merger rate and gas fraction are much higher for high
redshift galaxies, this evolutionary track must be an important channel for the formation
of massive red sequence galaxies over the cosmic time.

\acknowledgements We would like to thank the anonymous referee for several constructive reports
that improved the manuscript. We also thank Drs. Yu Gao,  Haojing Yan, Jiasheng Huang
and Sandy Faber for advice and helpful discussions. This project is supported by the NSF of China
11373027, 10973011, 11333003 and 11390372. It has also
been supported by the Strategic Priority Research Program ``The Emergence of
Cosmological Structures'' of the Chinese Academy of Sciences Grant No.
XDB09000000 (SM).
Funding for the creation and distribution of the SDSS Archive has been provided
by the Alfred P. Sloan Foundation, the Participating Institutions, the National
Aeronautics and Space Administration, the National Science Foundation, the U.S.
Department of Energy, the Japanese Monbukagakusho, and the Max Planck Society.
The SDSS Web site is http://www.sdss.org/.  The SDSS is managed by the
Astrophysical Research Consortium (ARC) for the Participating Institutions. The
Participating Institutions are The University of Chicago, Fermilab, the
Institute for Advanced Study, the Japan Participation Group, The Johns Hopkins
University, the Korean Scientist Group, Los Alamos National Laboratory, the
Max-Planck-Institute for Astronomy (MPIA), the Max-Planck-Institute for
Astrophysics (MPA), New Mexico State University, University of Pittsburgh,
Princeton University, the United States Naval Observatory, and the University
of Washington.  Some of the data presented in this paper were obtained from the
Mikulski Archive for Space Telescopes (MAST). STScI is operated by the
Association of Universities for Research in Astronomy, Inc., under NASA
contract NAS5-26555. Support for MAST for non-HST data is provided by the NASA
Office of Space Science via grant NNX09AF08G and by other grants and contracts.
This publication makes use of data products from the Wide-field Infrared
Survey Explorer, which is a joint project of the University of California, Los
Angeles, and the Jet Propulsion Laboratory/California Institute of Technology,
funded by the National Aeronautics and Space Administration.

\begin{deluxetable}{lcccc}
\tabletypesize{\scriptsize}
\tablecolumns{5}
\tablewidth{0pc}
\tablecaption{Physical Parameters for the Classification of ``Obscured Post-Starbursts''}
\tablehead{
\colhead{Source} & \colhead{EW(H$\alpha$)\tablenotemark{a}} &
\colhead{EW(H$\delta_A$)} & \colhead{$f_{12 \mu m}/f_{0.2 \mu m}$} &
\colhead{$f_{4.6 \mu m}/f_{3.4 \mu m}$} \\
\colhead{(1)} & \colhead{(2)} & \colhead{(3)} & \colhead{(4)} & \colhead{(5)}}
\startdata
  Q14168+0153 &  56.96$\pm$4.13  & 5.90$\pm$0.31 &  290.34$\pm$9.89   & 1.01$\pm$0.03 \\
  Q15257+0302 & 129.53$\pm$8.14  & 4.34$\pm$0.61 &  322.77$\pm$18.76  & 0.97$\pm$0.03 \\
  Q22364+1256 & 132.46$\pm$25.56 & 7.12$\pm$0.72 &  762.41$\pm$30.78  & 0.86$\pm$0.03 \\
  Q12383+2749 & 141.91$\pm$8.15  & 6.69$\pm$0.30 &  403.53$\pm$15.82  & 1.54$\pm$0.04 \\
  Q13293+0216 & 157.88$\pm$15.01 & 6.21$\pm$0.32 &  557.06$\pm$35.51  & 1.76$\pm$0.05 \\
  Q08550+3908 & 173.04$\pm$16.91 & 4.22$\pm$0.97 &  411.68$\pm$66.84  & 1.12$\pm$0.03 \\
  Q09343+6450 & 162.17$\pm$17.82 & 4.88$\pm$0.55 & 1674.52$\pm$143.05 & 0.86$\pm$0.02 \\
  Q15233+0533 &  88.54$\pm$8.51  & 6.95$\pm$0.38 &  377.91$\pm$24.92  & 0.88$\pm$0.02 \\
  F08559+1053 & 145.12$\pm$12.82 & 4.91$\pm$0.60 & 1142.33$\pm$136.60 & 1.47$\pm$0.04 \\
  F08591+5248 & 110.89$\pm$11.40 & 6.65$\pm$0.58 &  666.50$\pm$42.29  & 0.99$\pm$0.03 \\
  F09039+0503 & 138.04$\pm$13.94 & 4.13$\pm$0.78 &  262.73$\pm$15.15  & 1.02$\pm$0.04 \\
  F11387+4116 & 129.25$\pm$27.68 & 5.53$\pm$0.87 &  821.24$\pm$81.18  & 0.95$\pm$0.03 \\
\enddata
\tablenotetext{a}{Since the MPA/JHU catalog cautions the correctness of the H$\alpha$ continuum
uncertainties, we re-measured them to derive the uncertainties of the EW(H$\alpha$).}
\end{deluxetable}

\end{document}